\documentclass[prc,twocolumn,showpacs,preprintnumbers,amsmath,amssymb]{revtex4}
\usepackage{rotating}
\usepackage{graphicx}
\usepackage{dcolumn}
\usepackage{amstext}
\usepackage{lscape}
\usepackage{verbatim}
\usepackage[colorlinks,citecolor=blue,bookmarks]{hyperref}
\usepackage{times}
\usepackage{bm}
\usepackage{soul}
\begin{document}

\title{Tidal deformability of neutron and hyperon star with relativistic
mean field equations of state}
 
\author{Bharat Kumar$^{1,2}$}
\author{ S. K. Biswal$^{1,2}$}
%\author{Tanja Hinderer$^{3}$}
\author{S. K. Patra$^{1,2}$}

\affiliation{\it $^{1}$Institute of Physics, Bhubaneswar-751005, India} 
\affiliation{\it $^{2}$Homi Bhabha National Institute, Training School Complex, 
Anushakti Nagar, Mumbai 400085, India}
%\affiliation{\it $^{3}$Department of Physics, University of Maryland, College Park, Maryland 20742, USA}
\date{\today}

\begin{abstract}
We systematically study the tidal deformability for neutron and hyperon stars 
using relativistic mean field (RMF) equations of state (EOSs). The tidal 
effect plays an important role during the early part of the evolution of 
compact binaries. Although, the deformability associated with the EOSs has 
a small correction, it gives a clean gravitational wave signature in binary 
inspiral. These are characterized by various love numbers $k_l$ (l=2, 3, 4), 
that depend on the EOS of a star for a given mass and radius.  
The tidal effect of star could be efficiently measured 
through advanced LIGO detector from the final stages of inspiraling binary 
neutron star (BNS) merger.
\end{abstract}
\maketitle
%\footnotetext [1]{sbiswal@iopb.res.in}
\footnotetext [1]{bharat@iopb.res.in}
%\footnotetext [1]{patra@iopb.res.in}

\section{Introduction}\label{sec1}
The detection of gravitational waves is a major breakthrough in astrophysics/cosmology which is detected for the first time by 
advanced Laser Interferometer Gravitational-wave 
Observatory (aLIGO) detector \cite{black}.  Inspiraling and coalescing of 
binary black-hole results in emission of the gravitational waves.
We may expect that in a few years the forthcoming aLIGO \cite{black}, 
VIRGO \cite{virgo} and KAGRA \cite{kagra} detectors will also detect 
gravitational waves emitted by binary neutron star (NSs).
This detection will provide a valuable guidance and a better 
understanding of  highly compressed baryonic system. Flanagan and 
Hinderer \cite{flan,tanja,tanja1} have recently pointed out that tidal 
effects are also potentially measurable during the early part of the
evolution when the waveform is relatively clean. 
It is understood that the late inspiral signal may be
influenced by tidal interaction  between binary stars (NS-NS),
which gives the important information about the equation-of-state (EOS).
The study of Refs. \cite{baio,baio1,vines,pann,bd,dam,read,vines1,bd1,fava} inferred that the tidal effects could be measured using the recent generation of gravitational wave (GW) detectors.\\

In 1911, the famous mathematician A. E. H. Love \cite{love} introduced 
dimensional parameter in Newtonian theory that is related to the tidal 
deformation of the Earth is due to the gravitational attraction between the 
Moon and the Sun. These Newtonian theory of tides has been imported to 
the general relativity \cite{dam,pois}, where it  shows that the electric 
and magnetic type dimensionless gravitational Love number is a part of the 
tidal field  associated with the gravito-electric and gravitomagnetic 
interactions. The tidal interaction in a compact binary system has been 
encoded in the love number and is associated with the induced deformation
responded by changing shapes of the massive body. We are particularly 
interested for a neutron star in a close binary system, focusing on the 
various love number $k_l$ ($l$=2, 3, 4) due to the shape changes (like quadrupole,
octupole and hexadecapole in the presence of an  external gravitational field). 
Although higher love numbers ($l$=3, 4) give negligible effect, still these 
love numbers  ($k_{l}$) can have vital importance  in future gravitational wave 
astronomy.  However, geophysicists are 
interested to calculate the surficial love number $h_{l}$, which describe 
the deformation  of the body's surface in a multipole expansion \cite{dam,pois,land}. \\

We have used the equation of state from the relativistic mean filed 
(RMF) \cite{walecka74,rein89,ring96} and the newly developed effective
field theory motivated RMF (E-RMF) \cite{furn96,furn97} approximation 
in our present calculations. Here, the degrees of freedom are nucleon, 
$\sigma-,$ $\omega-$, $\rho-$, $\pi-$mesons and photon. 
This theory very well explains the properties 
of finite nuclei  and nuclear matter system at higher density region. 
Walecka has generalized \cite{walecka74} the RMF approximation and 
then subsequently Boguta and Bodmer \cite{boguta77} extended to 
the self-interaction of the $\sigma-$meson to reproduce proper experimental 
observables. In the E-RMF formalism, all the possible couplings of the mesons
among themselves and also their cross-interaction considered 
\cite{furn96,furn97}.  
The self- and crossed interactions are very significant. For example,
the self-interaction of $\sigma-$meson brings back the nuclear matter 
incompressibility $K_{\infty}$ from an unacceptable high value of 
$K_{\infty}\sim$600 MeV to a reasonable $K_{\infty}$ of $\sim$270 MeV 
\cite{boguta77,bodmer91}. Similarly, the quartic self-interaction of the 
vector meson $\omega$ soften the equation of state \cite{suga94,pieka05}.
It is to be noted that all the mesons and their  
self- and cross-interaction terms in the effective Lagrangian need not
to be included, because of some symmetry and their heavy masses \cite{mechl87}.
This theory of dense matter fairly explains the observed massive neutron star, 
like PSR J1614-2230 with mass $M=1.97\pm 0.04M_\odot$ \cite{demo10} and 
 PSR J0348+0432 ($M=2.01\pm 0.04 M_\odot$) \cite{antoni13}. 

The baryon octet are also introduced as
the stellar system is in extra-ordinary condition such as highly 
asymmetric or extremely high density medium \cite{bharat07}.
The coupling constants for nucleon-mesons are fitted to reproduce 
the properties of a finite number of nuclei, which then predict not only the
observables  of $\beta-$stable nuclei, but also of drip-lines and superheavy
regions \cite{walecka74,boguta77,rein86,ring90,rein88,sharma93,suga94,lala97}. 
The hyperon-meson couplings are obtained from the quark model \cite{oka,naka,naka1}. Recently, however, the couplings are improved by taking into consideration
some other properties of strange nuclear matter \cite{naka2}.

The paper is organized as follows: In Sec. \ref{form}, we give a brief 
description on relativistic mean field (RMF/E-RMF) formalism. The 
ingredients of the quantum hadrodynamic model (QHD) and resulting EOSs 
are outlined in this section. 
The various tidal love numbers and tidal deformability of neutron and 
hyperon star are discussed in Sec. \ref{love},\ref{tidal} after describing 
the numerical scheme. Finally, the summary and concluding remarks are 
given in Sec. \ref{conc}.

\section{The relativistic mean field formalism}\label{form}

The effective field theory motivated relativistic mean field (E-RMF)
Lagrangian is designed with underlying symmetries of quantum chromodynamics 
(QCD) and the parameters of G1 and G2 are constructed taking into account 
the naive dimensional analysis and naturalness 
 \cite{furn96,furn97,muller96,furn00,mach11}. For practical purpose, the terms
in the Lagrangian are taken upto $4^{th}$ order in 
meson-baryon couplings. The baryon-meson interaction is given by \cite{bharat07}:   

\begin{eqnarray}
{\cal L}&=&\sum_B\overline{\psi}_B\left(
i\gamma^{\mu}D_{\mu}-m_B+g_{\sigma B}\sigma%+g_{\delta B}\delta.\tau 
\right)\psi_B \nonumber \\
&& + \frac{1}{2}\partial_{\mu}\sigma\partial_{\mu}\sigma-m_{\sigma}^2\sigma^2
\left(\frac{1}{2}+\frac{g_3}{3!}\frac{g_{\sigma}\sigma}{m_B}
+\frac{g_4}{4!}\frac{g_{\sigma}^2\sigma^2}{m_B^2}\right) \nonumber \\
&& - \frac{1}{4}\Omega_{\mu\nu}\Omega^{\mu\nu}+\frac{1}{2}m_{\omega}^2
\omega_{\mu}\omega^{\mu}\left(1+\eta_1\frac{g_{\sigma}\sigma}{m_B}
+\frac{\eta_2}{2}\frac{g_{\sigma}^2\sigma^2}{m_B^2}\right) \nonumber \\
&& -\frac{1}{4}R_{\mu\nu}^aR^{\mu\nu a}+\frac{1}{2}m_{\rho}^2
\rho_{\mu}^a\rho^{a\mu}\left(1+\eta_{\rho}
\frac{g_{\sigma}\sigma}{m_B} \right) \nonumber \\
&&%+\frac{1}{2}\partial_{\mu}\delta.\partial_{\mu}\delta-m_{\delta}^2\delta^2
+\frac{1}{4!}\zeta_0 \left(g_{\omega}\omega_{\mu}\omega^{\mu}\right)^2
+\Lambda_v(g_{\rho}^2\rho_{\mu}^a\rho^{\mu a})(g_{\omega}^{2}\omega_{\mu}\omega^{\mu})\nonumber \\
&&+\sum_l\overline{\psi}_l\left(
i\gamma^{\mu}\partial_{\mu}-m_l\right)\psi_l.%\nonumber
%+\Lambda_v(g_{\rho}^2\rho_{\mu}^a\rho^{\mu a})
%(g_{\omega}^2\omega_{\mu}\omega^{\mu}).
\end{eqnarray}
The co-variant derivative $D_{\mu}$ is defined as:
\begin{eqnarray}
D_{\mu}=\partial_{\mu}+ig_{\omega}\omega_{\mu}+ig_{\rho}I_3\tau^a\rho_{\mu}^a,
\end{eqnarray}
where $R_{\mu\nu}^a$ and $\Omega_{\mu\nu}$ are field tensors defined as:
\begin{eqnarray}
R_{\mu\nu}^a=\partial_{\mu}\rho_{\nu}^a-\partial_{\nu}\rho_{\mu}^a
+g_{\rho}\mathcal{E}_{abc}\rho_{\mu}^b\rho_{\nu}^c,
\end{eqnarray}
\begin{eqnarray}
\Omega_{\mu\nu}=\partial_{\mu}\omega_{\nu}-\partial_{\nu}\omega_{\mu}.
\end{eqnarray}

Here, the symbol B stands for the baryon octet (n,p,$\Lambda$, $\Sigma^{+}$,
$\Sigma^{0}$,$\Sigma^{-}$,$\Xi^{-}$,$\Xi^{0}$) and $l$ represents 
e$^{-}$ and $\mu^{-}$. The masses $m_{B}$, $m_{\sigma}$, $m_{\omega}$, and 
$m_{\rho}$ are, respectively for baryons and for  $\sigma$, $\omega$ and 
$\rho$ meson fields. In real calculation, we ignore the non-abelian 
term from the $\rho-$field.  $I_3$ is the third component of isospin projection. All symbols are carrying their own usual meaning~ \cite{bharat07,sing14}.

For a  given Lagrangian density in Eq.(1), one can solve the 
equations of motion \cite{bharat07,sing14} in the mean-field level, i.e. the
exchange of mesons create an uniform meson field, where the nucleon has
a simple harmonic motion. Then we calculate the energy-momentum tensor
within the mean field approximation (i.e. the meson fields are replaced
by their classical number) and get the EOS as a function of baryon density. 
 The EOS remains uncertain at density larger than the saturation density of 
nuclear matter, $\rho_{n}\sim$3 $\times$ 10$^{14}$ g cm$^{-3}$.
 At these densities, neutrons can no longer be considered, 
which may  consists mainly of heavy
baryons (mass greater than nucleon) and several other species of particles 
expected to appear due to the rapid rise of the baryon chemical potentials
 \cite{baldo}. 
The $\beta$-equilibrium and  charge neutrality  
are two important conditions for the neutron/hyperon rich-matter. 
Both these conditions force the stars to  have $\sim$90$\%$ of neutron 
and $\sim$10$\%$ proton. With the inclusion of  baryons, the 
$\beta-$equilibrium conditions between chemical potentials  for different 
particles: 
\begin{eqnarray}\label{beta}
\mu_p = \mu_{\Sigma^+}=\mu_n-\mu_{e}, \nonumber\\
\mu_n=\mu_{\Sigma^0}=\mu_{\Xi^0}=\mu_{n},\nonumber \\
\mu_{\Sigma^-}=\mu_{\Xi^-}=\mu_n+\mu_{e},\nonumber \\
\mu_{\mu}=\mu_{e}, \nonumber\\
\end{eqnarray}
and the charge neutrality condition is satisfy by  
\begin{eqnarray}\label{charge}
n_p+n_{\Sigma^+}=n_e+n_{\mu^-}+n_{\Sigma^-}+n_{\Xi^-}.
\end{eqnarray}
The corresponding pressure and energy density of the charge neutral 
beta-equilibrated neutron star matter (which includes the lowest lying octet of 
baryons) is then given by \cite{bharat07}:

\begin{eqnarray}\label{energy}
\mathcal{E}&=&\sum_B\frac{2}{(2\pi)^3}\int_0^{k_B}d^3k E_B^*(k)
+\frac{1}{8}\zeta_0g_{\omega}^2\omega_0^4 \nonumber \\ 
&& + m_{\sigma}^2\sigma_0^2\left(\frac{1}{2}+\frac{g_3}{3!}
\frac{g_{\sigma}\sigma_0}{m_B}+\frac{g_4}{4!}
\frac{g_{\sigma}^2\sigma_0^2}{m_B^2}\right) \nonumber \\
&& + \frac{1}{2}m_{\omega}^2 \omega_0^2\left(1+\eta_1
\frac{g_{\sigma}\sigma_0}{m_B}+\frac{\eta_2}{2}
\frac{g_{\sigma}^2\sigma_0^2}{m_B^2}\right) \nonumber \\
&& + \frac{1}{2}m_{\rho}^2 \rho_{03}^2\left(1+\eta_{\rho}
+ \frac{g_{\sigma}\sigma_0}{m_B} \right) 
+3 \Lambda_v(g_{\rho}\rho_{03})^{2}(g_{\omega}\omega_{0})^{2}
\nonumber \\ &&
%+\frac{1}{2}\frac{m_{\delta}^2}{g_{\delta}^{2}}(\delta^3)^2
+\sum_l\mathcal{E}_l,
\end{eqnarray}

and

\begin{eqnarray}\label{pressure}
P&=&\sum_B\frac{2}{3(2\pi)^3}\int_0^{k_B}d^3k \frac{k^{2}}{E_B^*(k)}
+\frac{1}{8}\zeta_0g_{\omega}^2\omega_0^4 \nonumber \\ 
&& - m_{\sigma}^2\sigma_0^2\left(\frac{1}{2}+\frac{g_3}{3!}
\frac{g_{\sigma}\sigma_0}{m_B}+\frac{g_4}{4!}
\frac{g_{\sigma}^2\sigma_0^2}{m_B^2}\right) \nonumber \\
&& + \frac{1}{2}m_{\omega}^2 \omega_0^2\left(1+\eta_1
\frac{g_{\sigma}\sigma_0}{m_B}+\frac{\eta_2}{2}
\frac{g_{\sigma}^2\sigma_0^2}{m_B^2}\right) \nonumber \\
&& + \frac{1}{2}m_{\rho}^2 \rho_{03}^2\left(1+\eta_{\rho}
\frac{g_{\sigma}\sigma_0}{m_B} \right)
+ \Lambda_v(g_{\rho}\rho_{03})^{2}(g_{\omega}\omega_{0})^{2} 
\nonumber \\&&%-\frac{1}{2}\frac{m_{\delta}^2}{g_{\delta}^{2}}(\delta^3)^2
+\sum_l P_l,
\end{eqnarray}
where $P_l$ and $\mathcal{E}_l$ are lepton's pressure and energy 
density, respectively. 
 $E_B^*=(k_B^2+M_B^{*2})^{1/2}$ is the effective energy, $k_B$ is
the Fermi momentum of the baryons. $M_p^*$ and $M_n^*$ are the proton and
neutron effective masses written as\\
\begin{eqnarray}
M_p^*=M_p-g_{\sigma}\sigma_0%-g_{\delta}{\delta^3}\\
\end{eqnarray} 
and 
\begin{eqnarray}
M_n^*=M_n-g_{\sigma}\sigma_0.%+g_{\delta}{\delta^3},
\end{eqnarray}

\begin{figure}[ht]
\includegraphics[width=1.0\columnwidth]{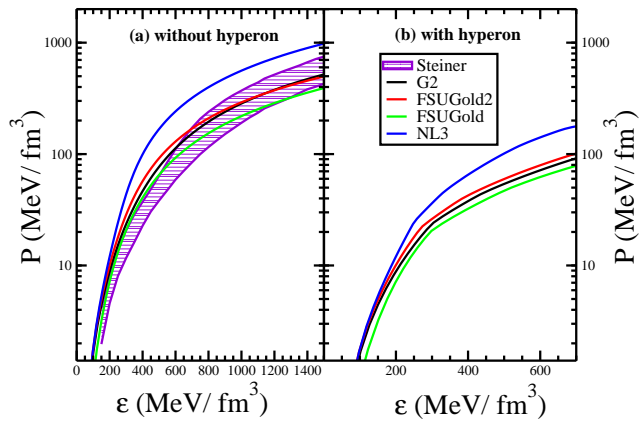}
\caption{(Color online) The equations of state obtained for nuclear and 
hyper-nuclear matter under charge neutrality as well as the $\beta-$ 
equilibrium condition for G2 \cite{furn97}, FSUGold2 \cite{fsu2}, 
 FSUGold \cite{pieka05} and NL3 \cite{lala97} force parameters are compared 
with the empirical data \cite{ste10} (shaded area in the graph)for r$_{ph}$=R 
with the uncertainty of $2\sigma$. Here, R and r$_{ph}$ are the neutron radius 
and the photospheric radius, respectively. }
\label{fig1}
\end{figure}

\section{Results and Discussions:}

\subsection{Equations of state}

In this section, we present the results of our calculations in Figs. 1-9 and
Table \ref{tab2}. Before going to the estimation of the tidal deformibility parameter
$\lambda$, we check the validity of the EOSs obtained with various force
parameters.  Fig.\ref{fig1} displays the equation of state for G2 \cite{furn97},
FSUGold2 \cite{fsu2}, FSUGold \cite{pieka05} and NL3 \cite{lala97} parameter
sets. From the left panel of Fig. \ref{fig1}(a), it is obvious that all the EOSs 
follow similar trend. Among these four, the celebrity NL3 set gives the 
stiffest EOS and the relatively new FSUGold represents the softer character.
This is because of the large and positive $g_{4}$ value as well as 
the introduction of isoscalar-isovector coupling ($\Lambda$) in the FSUGold 
set \cite{pieka05}.  To have an understanding on the softer and stiffer EOSs by 
various parametrizations, we compared their coupling constants and other 
parameters of the sets in Table \ref{tab1}. We notice a large 
variation in their effective masses, incompressibilities and other
nuclear matter properties at saturation. 
For higher  energy density $\mathcal{E} \sim 500-1400$ MeV fm$^{-3}$, except 
NL3 set, which has the lowest nucleon effective mass, all other sets are found 
in the region of empirical  data with the uncertainty of 2$\sigma$ \cite{ste10}.

Fig. \ref{fig1}(b) shows a hump type structure on the
nucleon-hyperon equation of state at $\mathcal{E}$ around 400-500 MeV 
fm$^{-3}$.  This kink ($\mathcal{E}\sim$ 200-300 MeV) shows
the presence of hyperons in the dense system. Here, the repulsive component
of the vector potential becomes more important than the attractive part
of the scalar interaction.  As a result the coupling of the 
hyperon-nucleon strength gets weak. 
At a given baryon density, the inclusion of hyperons lower  significantly the  
pressure compared to the equation of state of having without hyperons. This is 
possible due to the higher energy of the hyperons, as the neutrons are replaced by the low-energy hyperons.  The hyperon couplings are expressed as the 
ratio of the meson-hyperon and meson-nucleon couplings:
\begin{eqnarray}
\chi_{\sigma} = \frac {g_{Y\sigma}}{g_{N\sigma}}, \chi_{\omega} = \frac {g_{Y\omega}}{g_{N\omega}},\chi_{\rho} = \frac {g_{Y\rho}}{g_{N\rho}}.
\end{eqnarray}
In the present calculations, we have taken $\chi_{\sigma}$ = 0.7 and 
$\chi_{\omega}=\chi_{\rho} $ = 0.783. One can
find similar calculations for stellar mass in Refs. \cite{glen20,mene,weis12,lope14}.

\begin{table}
\hspace{0.1 cm}
\caption{Parameters and saturation properties for NL3 \cite{lala97}, 
G2 \cite{furn97}, FSUGold \cite{pieka05}, and FSUGold2 \cite{fsu2}. The 
parameters $g_\sigma$, $g_\omega$, $g_\rho$, $g_{3}$, and $g_{4}$ are 
calculated from nuclear matter the given saturation properties using 
the relations suggested by the authors of Ref. \cite {glen20}. }
\renewcommand{\tabcolsep}{0.1 cm}
\renewcommand{\arraystretch}{1.2}
{\begin{tabular}{|c|c|c|c|c|}
\hline
\cline{1-5}
\hline
Parameters &       NL3   &      G2 &FSUGold  &     FSUGold2 \\
\hline
&&&&\\
$m_{n}$(MeV)&939&939&939&939\\
$m_\sigma$(MeV)&508.194&520.206&491.5&497.479\\
$m_\omega$(MeV)&782.501&782&783&782.5\\
$m_\rho$(MeV)&763&770&763&763\\
$g_\sigma$&10.1756&10.5088&10.5924&10.3968\\
$g_\omega$&12.7885&12.7864&14.3020&13.5568\\
$g_\rho$&8.9849&9.5108&11.7673&8.970\\
$g_{3}$(MeV)&1.4841&3.2376&0.6194&1.2315\\
$g_{4}$&-5.6596&0.6939&9.7466&-0.2052\\
$\eta_{1}$&0&0.65&0&0\\
$\eta_{2}$&0&0.11&0&0\\
$\eta_\rho$&0&0.390&0&0\\
$\zeta_{0}$&0&2.642&12.273&4.705\\
$\Lambda$&0&0&0.03&0.000823\\
$\rho_{0}$(fm$^{-3})$&0.148&0.153&0.148&0.1505$\pm$0.00078\\
E/A(MeV)&-16.299&-16.07&-16.3&-16.28$\pm$0.02\\
K$_\infty$(MeV)&271.76&215&230&238.0$\pm$ 2.8\\
J(MeV)&37.4&36.4&32.59&37.62$\pm$1.11\\
L(MeV)&118.2&101.2&60.5&112.8$\pm$ 16.1\\
$m_{n}^*$/$m_{n}$&0.6&0.664&0.610&0.593$\pm$0.004\\

\hline
\end{tabular}\label{tab1} }
\end{table}

\subsection{Mass and radius of neutron star}

\begin{figure}[ht]
\includegraphics[width=1.0\columnwidth]{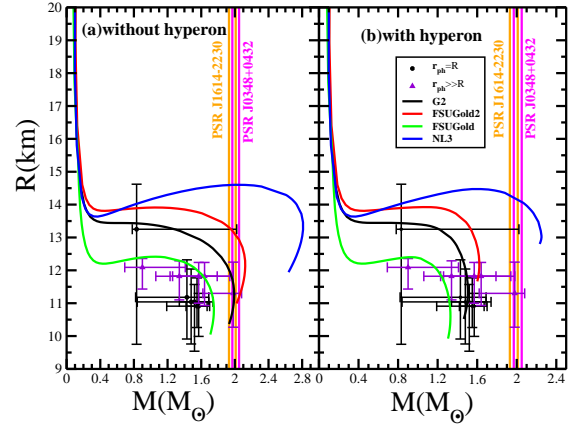}
\caption{(Color online)  The mass-radius profile for the force parameters
 like G2 \cite{furn97}, FSUGold2 \cite{fsu2}, FSUGold \cite{pieka05} and
NL3 \cite{lala97} used. The solid circles($r_{ph}$=R) and triangles($r\gg$R)
 are  represent the observational constraints \cite{ste10}, where $r_{ph}$ is
the photospheric radius. The verticle sheded region correspond to the recent
observation \cite{demo10,antoni13}.}
\label{fig2}
\end{figure}

Once the equations of state for various relativistic forces are fixed, then
we extend our study for the evaluation of the mass and radius of the isolated
neutron star. The Tolmann-Oppenheimer-Volkov (TOV) equations \cite{tolm39} 
have to be solved for this purpose, where EOSs are the inputs. The TOV equations
are written as:
\begin{eqnarray}
\frac{d P(r)}{d r}=-\frac{[{\mathcal{E}}(r)+P(r)][M(r)+{4\pi r^3 P(r)}]}{r^2(1-\frac{2M(r)}{ r})}, 
\end{eqnarray} 
and
\begin{eqnarray}
\frac{d M(r)}{d r}={4\pi r^2 {\mathcal{E}}(r)}.
\end{eqnarray}
For a given EOS, the Tolmann-Oppenheimer-Volkov (TOV)
equations \cite{tolm39} must be integrated from the boundary
conditions $P(0) = P_{c}$, and $M(0)$=0, where $P_{c}$ and M(0) are 
the pressure and mass of the star at $r$=0 and the value of $r$(= $R$), 
where the pressure vanish defines the surface of the star. 
Thus, at each central density we can uniquely determine a mass $M$ and 
a radius R  of the static neutron and  hyperon satrs using the four chosen 
EOSs. The estimated result for the maximum mass as a function of radius 
are compared with the highly precise measurements of two massive 
($\sim 2 M_\odot$) neutron star \cite{demo10,antoni13} and extraction of 
stellar radii from X-ray observation \cite{ste10}, are shown in Figs. 
\ref{fig2}(a) and \ref{fig2} (b) . From recent observations 
 \cite{demo10,antoni13}, it is clearly illustrated that the maximum mass 
predicted by any theoretical models should reach the limit $\sim 2.0 M_\odot$, 
which is consistent with our present prediction from the G2 equation of state
 of nucleonic matter compact star with mass 1.99$M_\odot$ and
radius 11.25 km.   From X-ray observation, Steiner  et. al. \cite{ste10} 
predicted that the most-probable neutron star radii lie in the range 11-12 km 
with neutron star masses$\sim$1.4$M_\odot$ and predicted the EOS is relatively 
soft in the density range 1-3 times the nuclear saturation density. As 
explained to earlier, stiff EOS like NL3 predicts larger stellar radius 
13.23 km and  a maximum mass 2.81 $M_\odot$.  Though  FSUGold and 
FSUGold2 are from the same RMF model with similar terms in the Lagrangian, 
their results for neutron star are quite different with FSUGold2 suggesting 
a larger  and heavier NS with mass  2.12$M_\odot$ and radius 12.12 km 
compare to mass and radius (1.75$M_\odot$ and 10.76 km) of the FSUGold. 
Because in FSUGold2 EOS at high densities, the impact comes from the quartic 
vector  coupling constant $\zeta_{0}$ and also the large value of the slope
parameter L=112.8 $\pm$16.1 MeV (see Table \ref{tab1}) tend to predict the 
neutron star with large radius \cite{horo01}. 
 From the observational point of view, there are large uncertainties in the 
determination of the radius of the star \cite{rute,gend,cott}, which is 
a hindrance to get a precise knowledge on the composition of the star 
atmosphere. One can see that G2 parameter is able to reproduce the recent 
observation of 2.0$M_\odot$ NS. But the presence of hyperon matter under 
$\beta$-equilibrium soften the EOS, because they are more massive
than nucleons and when they start to fill their Fermi sea slowly 
replacing the highest energy nucleons. Hence, the maximum mass of NS is 
reduced by $\sim$0.5 unit solar mass due to the high baryon density. 
For example, the stiffer NL3 equation of state gives the
maximum NS mass $\sim$2.81$M_\odot$ and the presence of hyperon-matter reduces
the mass to $\sim$2.25$M_\odot$ as shown in Fig. \ref{fig2}(b).
These results give us warning that most of the present sets of hyperon 
couplings unable to reproduce the recently observed mass of neutron star 
like PSR J1614-2230 with mass  $M = 1.97\pm 0.04M_{\odot}$ \cite{demo10} and 
the PSR J0348+0432 with  $M = 2.01\pm 0.04M_{\odot}$ \cite{antoni13}. 
Probably, this suggest us to modify the coupling constants
and get the equations of state proper, so that one can explain all the
mass-radius observation till date. Further, one can see that in 
Fig. \ref{fig2}(b) mass-radius curve of G2, FSUGold, FSUGold2 with hyperon 
lies in  the range of predicted equation of state between the
$r_{ph}$=R and $r_{ph}\gg$R cases is the high density behaviour \cite{ste10}.

\subsection{ Various tidal love number of compact star}\label{love}

When spherical star  placed in a static external quadrupolar tidal
field $\mathcal{E}_{ij}$ then the star will  be deformed and quadrupole
deformation will be the leading order perturbation. Such a deformation is
defined as the ratio of the mass quadrupole moment of a star Q$_{ij}$ to 
the external tidal field $\mathcal{E}_{ij}$:
\begin{eqnarray}
\lambda= -\frac{Q_{ij}}{\mathcal{E}_{ij}}.
\end{eqnarray}
Specifically, the observable of the tidal deformability parameter
$\lambda$ depends on the EOS via both the neutron star (NS) radius and a 
dimensionless quantity k$_{2}$, called the Love number and is given by the 
relation:
\begin{eqnarray}
\lambda = \frac{2}{3} k_{2}R^{5},
\end{eqnarray}
and the dimensionless tidal-deformability($\Lambda$) is related with the 
compactness parameter $C=M/R$ as:
\begin{eqnarray}
\Lambda = \frac{2 k_{2}}{3 C^{5}}, 
\end{eqnarray}
where R is the radius of the (spherical) star in isolation.
Now, we have to get k$_{2}$ for the calculation of the deformability parameter
 $\lambda$, which is the key quantity of deformation due to the gravitational
attraction of the binary stars with each other. This force of attraction becomes
more and more important in the course of time, because of the reduction of the
orbital distance between them. The orbital distance between the binary decreases
as the companion star emits gravitational radiation. As a result, the binary
accelerates and finally merge with each other and possibly turns to a black
hole. Thus, the estimation of the leading order quadrupole electric tidal 
love number $k_{2}$ along with other higher order love numbers $k_{3}$ 
and $k_{4}$ are very important for the detection of gravitational wave. 

To estimate the love numbers $k_l$ ($l$=2, 3, 4), along with the evaluation
of the TOV equations, we have to compute $y = y_{l}(R)$ with initial
boundary condition $y(0)=l$ from the following 
first order differential equation iteratively \cite{tanja,tanja1,thib1,pois}: 
\begin{eqnarray}
r\frac{d y(r)}{d r} + y(r)^{2} + y(r) F(r) + r^{2} Q(r)=0,
\end{eqnarray}
with,
\begin{eqnarray}
 F(r)=\frac{r - 4 \pi r^{3} [{\mathcal{E}}(r)-P(r)]}{r-2 M(r)},\\
 Q(r)=\frac{4 \pi r(5{\mathcal{E}}(r)+9 P(r)+\frac{{\mathcal{E}}(r)+P(r)} 
{{\partial P(r)}/ {\partial{\mathcal{E}}(r)}}-\frac{l (l+1)}{4 \pi r^{2}})}{r-2 M(r)}
\nonumber \\
-4 \Big[\frac {M(r)+4 \pi r^{3} P(r)}{r^{2}(1- 2 M(r)/r)}\Big]^{2}.
\end{eqnarray}
Once, we know the value of $y= y_{l}(R)$, the electric tidal love numbers 
$k_{l}$ are found from the following expression \cite{thib1}:

\begin{widetext}
\begin{eqnarray}
k_{2} = \frac{8}{5}(1-2C)^{2}C^{5}[2C(y_{2}-1)-y_{2}+2]\Big\{ 2C(4(y_{2}+1)C^{4}
+(6y_{2}-4)C^{3}+
\nonumber \\
(26-22y_{2})C^{2}+3(5y_{2}-8)C-3y_{2}+6)
-3(1-2C)^{2}(2C(y_{2}-1)-y_{2}+2)log\Big( \frac{1}{1-2C}\Big)\Big\}^{-1},   
\end{eqnarray}
%\end{widetext}

%\begin{widetext}
\begin{eqnarray}
k_{3} = \frac{8}{7}(1-2C)^{2}C^{7}[2(y_{3}-1)C^{2}-3(y_{3}-2)C+y_{3}-3]
\Big\{  2C[4(y_{3}+1) C^{5}+2(9y_{3}-2)C^{4}
\nonumber \\ 
-20(7y_{3}-9)C^{3}+5(37y_{3}-72)C^{2}
-45(2y_{3}-5)C+15(y_{3}-3)]-15(1-2C)^{2}
\nonumber \\
(2(y_{3}-1)C^{2}-3(y_{3}-2)C+y_{3}-3)log\Big( \frac{1}{1-2C}\Big)\Big\}^{-1},
\end{eqnarray}
%\end{widetext}
and
%\begin{widetext}
\begin{eqnarray}
k_{4} = \frac{32}{147}(1-2C)^{2}C^{9}[12(y_{4}-1)C^{3}-34(y_{4}-2)C^{2}
+28(y_{4}-3)C-7(y_{4}-4)]\Big\{ 2C[8(y_{4}+1)C^{6}
\nonumber \\
+(68y_{4}-8)C^{5}+(1284-996y_{4})C^{4}+40(55y_{4}-116)C^{3}+(5360-1910y_{4})
C^{2}
+105(7y_{4}-24)C-105(y_{4}-4)]
\nonumber \\
-15(1-2C)^{2}[12(y_{4}-1)C^{3}-34(y_{4}-2)C^{2}+28(y_{4}-3)C-7(y_{4}-4)] log\Big(\frac{1}{1-2C}\Big)\Big\}^{-1} ,
\end{eqnarray}
\end{widetext}

\begin{figure}[ht]
\includegraphics[width=1.0\columnwidth]{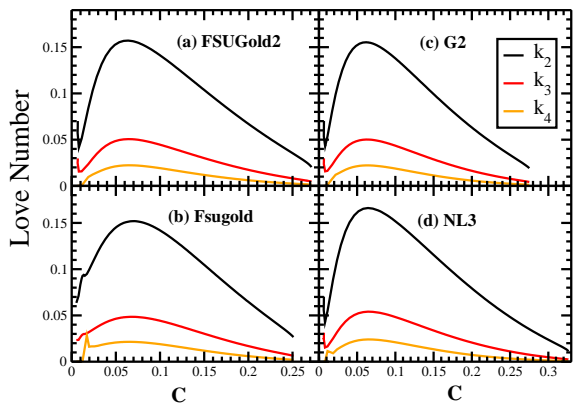}
\caption{(Color online)The tidal love numbers $k_{2},k_{3},k_{4}$ as a
function of the mass of the four selected EOSs of the neutron star.}
\label{fig3}
\end{figure}
\begin{figure}[ht]
\includegraphics[width=1.0\columnwidth]{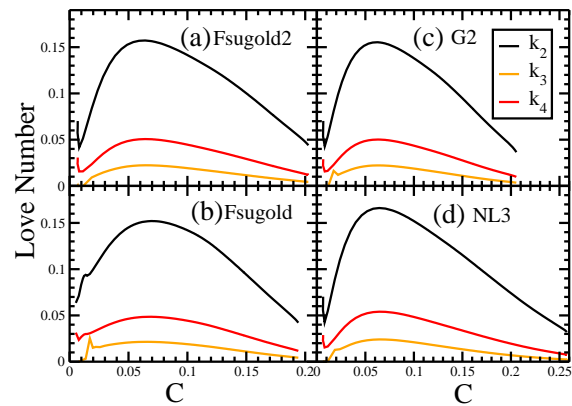}
\caption{(Color online) Same as Fig. \ref{fig3} but for hyperon star.}
\label{fig4}
\end{figure}
As we have emphasized earlier, the dimensionless love number $k_{l}$ 
(l=2, 3, 4) is an important quantity to
measure the internal structure of the constituent body. These quantities
directly enter into  the gravitational wave phase of inspiralling
binary neutron star (BNS) and extract the information of the EOS. 
Notice that equations (20)-(22) contain an overall factor $(1- 2C)^{2}$,
which tends to zero when the compactness approaches the compactness 
of the black hole, i.e. $C^{BH}$=1/2 \cite{thib}. 
Also, it is to be pointed out that the presence of 
multiplication order factor $C$ with $(1- 2C)^{2}$ in the expression
of $k_l$ that the value of the love number of a black hole simply becomes
zero, i.e. $k^{BH}_{l}$=0.

\begin{table*}
\hspace{0.1 cm}
\caption{Properties of a 1.4$M_\odot$ neutron  and hyperon star for 
different  class of the EOS. The quadrupolar tidal polarizability $\lambda$ 
and uncertainty error $\Delta\tilde{\lambda}$ in (10$^{36}$g cm$^{2}$ s$^{2}$).}
\renewcommand{\tabcolsep}{0.1 cm}
\renewcommand{\arraystretch}{1.2}
{\begin{tabular}{|c|c|c|c|c|c|c|c|c|c|c|c|c|c|c|c|c|c|c|c|c|c|c|}
\hline
 \multicolumn{13}{ |c| }{Neutron Star}\\
\cline{2-13}%\cline{6-7}\cline{7-10}
\hline
EOS &       $R(km)$   &       $C$ &$f_{c}$(Hz)  &       $k_{2}$       &       $k_{3}$        &       $k_{4}$       &  $h_{2}$ & $h_{3}$ &$h_{4}$&     $\lambda$
&$\Delta\tilde{\lambda}$&$\Lambda$\\
\hline
&&&&&&&&&&&&\\
NL3&14.422&0.144&1256.7&0.1197&0.0353&0.0142&0.9775&0.6519&0.5074&7.466&2.027
&1288.81\\
G2&13.148&0.157&1440.9&0.0934&0.0265&0.0103&0.8879&0.5951&0.4596&3.668&1.486
&652.76\\
FSUGold2&13.850&0.149&1332.4&0.1040&0.0301&0.0119&0.9275&0.6237&0.4854&5.299&1.763&944.08\\
FSUGold&12.236&0.170&1608.0&0.0882&0.0244&0.0071&0.8589&0.5634&0.4268&2.418&1.178&414.13\\
\hline
\multicolumn{13}{ |c| }{Hyperon Star}\\
\cline{1-13}
&&&&&&&&&&&&\\
NL3&14.430&0.143&1252.9&0.1203&0.0355&0.0143&0.9800&0.6541&0.5096&7.527&2.018&1341.20\\
G2&12.686&0.163&1520.6&0.0804&0.0229&0.0088&0.8434&0.5707&0.4399&2.641&1.321&465.83\\
FSUGold2&13.690&0.151&1355.9&0.0988&0.0287&0.0113&0.9108&0.6154&0.4789&4.750&1.696&839.04\\
(FSUGold)$_{1.3M_\odot}$&9.922&0.194&2119.0&0.0421&0.0116&0.0042&0.6884&0.4683&0.3518&0.4048&0.530
&102.14\\
\hline
\end{tabular}\label{tab2} }
\end{table*}

Fig. \ref{fig3} shows the tidal love
numbers $k_{l}$ ($l$=2, 3, 4) as a function compactness parameter $C$ 
for the neutron star with four selected EOSs. The result of $k_{l}$ 
suddenly deceases with increasing compactness (C = 0.06-0.25). 
For each EOS, the value of $k_{2}$ appears to be a maximum between 
$C=0.06-0.07$. However, we are mainly interested in the neutron star 
masses at $\sim$1.4$M_\odot$.  Because of the tidal interactions in the 
neutron star binary, the shape of the star acquires quadrupole,
octupole, hexadecapole and other higher order deformations. The value
of the love numbers for corresponding shapes are shown in Table \ref{tab2}. 
The values of $k_l$ decreases gradually with increase of multi-pole moments. 
Thus, the quadrupole deformibility has the maximum effects on the
binary star merger.   
Similarly, in Fig. \ref{fig4}, the dimensionless love number $k_{l}$ is shown as a 
function of compactness for the hyperon star. With the inclusion of 
hyperons, the effect of the core is negligible due to the softness of the 
EOSs. The values of $k_{l}$ is different for a typical  neutron-hyperon star 
with 1.4 $M_{\odot}$ for various sets are listed in the lower portion of
Table \ref{tab2}. The radius and respective mass-radius ratio is also given in 
the Table \ref{tab2}. The table also reflects that the love numbers  
decrease  slightly or remains unchanged with the addition of hyperon in the 
neutron star. 
 The neutron star surface or solid crust is not responsible for any tidal
effects, but instead it is the matter mainly in the outer core that gives
the largest contribution to the tidal love numbers. It is relatively 
unaffected by changing the composition of the core and leave it at that.  
Thus instigate the calculation for the surficial  love number
$h_l$ for both neutron and hyperon star binary. 
  
\begin{figure}[ht]
\includegraphics[width=1.0\columnwidth]{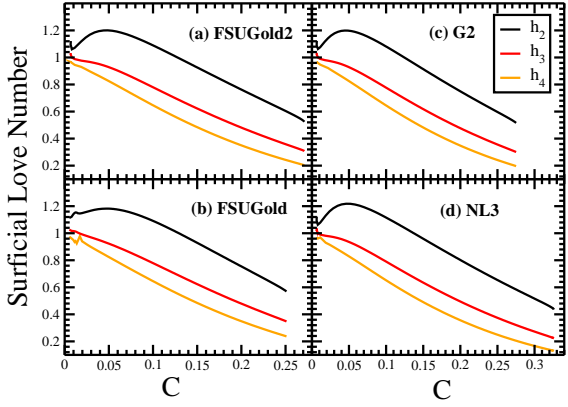}
\caption{(Color online) Surficial love number h$_{l}$ as a function
of compactness C of a neutron star, for selected values of $l$.}
\label{fig5}
\end{figure}
\begin{figure}[ht]
\includegraphics[width=1.0\columnwidth]{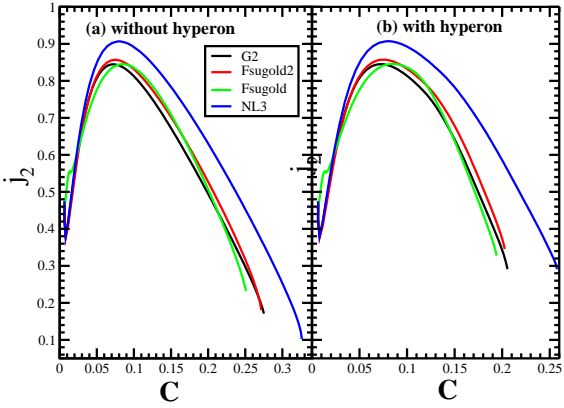}
\caption{(Color online) The magnetic tidal love number for
selected EOSs. }
\label{fig6}
\end{figure}
Next, we calculate the surficial love number $h_{l}$ which describes the 
deformation of the body's surface in a multipole expansion. Recently, Damour 
and Nagar \cite{thib} have given the surficial love number
(also known as shape love number) $h_{l}$ for the coordinate displacement 
$\delta R$ of the body's surface under the external tidal force. 
Alternatively, Landry and Poisson \cite{land} have proposed the defination 
of Newtonian love number in terms of a curvature perturbation $\delta \mathcal{R}$ instead of a surface displacement $\delta R$. 
For a perfect fluid, the relation between the surficial love number $h_{l}$ 
and tidal love number $k_{l}$ is given as
\begin{eqnarray}
  h_{l}=\Gamma_{1}+2 \Gamma_{2} k_{l}\\
 \Gamma_{1}=\frac{l+1}{l-1}(1-C)F(-l,-l,-2l;2C)\nonumber\\
-\frac{2}{l-1}F(-l,-l-1,-2l;2C),\nonumber\\  
\Gamma_{2}=\frac{l}{l+2}(1-C)F(l+1,l+1,2l+2;2C)\nonumber\\
+\frac{2}{l+2}F(l+1,l,2l+2;2C).
\end{eqnarray}
where F(a,b,c;z) is the hypergeometric function.
Fig. \ref{fig5}  shows the results of surficial love number $h_l$
of a neutron star as a function of compactness parameter C. Unlike  
the initially increasing and then decreasing trend of the tidal love 
number $k_l$, the surficial love number $h_l$ decreases almost 
exponentially with the compactness parameter. 
At the minimum value of the compactness parameter, the maximum value of the
shape love number of each multipole moment approaches 1.
Thus, we zero in on to the Newtonian relation i.e $h_{l}=1+2k_{l}$.
Again one can compute from Table \ref{tab2} 
that the surficial love number $h_l$ decreases $\sim 20 \%$ from one moment 
to another. For example,  $h_2=0.9775$ and $h_3=0.6519$ and $h_4=0.5074$ 
for NL3 parameter sets.

Furthermore,  we also calculate the "magnetic" tidal love number $j_l$. Here,
we give only the quadrupolar case ($l=2$), which is expressed as: 
\begin{eqnarray}
j_{2}=\Big \{96 C^{5}(2C-1)(y_{2}-3)\Big\}\Big\{5(2C(12(y_{2}+1)C^{4}
\nonumber \\+2(y_{2}-3)C^{3}+2(y_{2}-3)C^{2}+3(y_{2}-3)C-3y_{2}+9)
\nonumber \\+3(2C-1)(y_{2}-3)log(1-2C))\Big\}^{-1}. 
\end{eqnarray}
After inserting the value of $y_{2}$ in eq. (25), we compute the  magnetic 
tidal love number $j_{2}$ in a hydrostatic equilibrium condition for a 
non-rotating neutron star.  This gives important information about the internal 
structure \cite{pois} without changing the  tidal love number $k_{2}$.  
At $C$=0.01, the magnetic love number $j_{2}$ is nearly 0.4. In both cases
(with and without hyperons), $j_{2}$ is maximum within the compactness 
0.06 to 0.07 for all the four EOSs (See Fig. \ref{fig6}). Then the value of 
$j_2$ the decreases sharply with increase of compactness. The NL3 parameter 
set gives a maximum $j_{2}$ in both the systems, while rest of the three sets
predict comparable $j_2$.

\begin{figure}[ht]
\includegraphics[width=1.0\columnwidth]{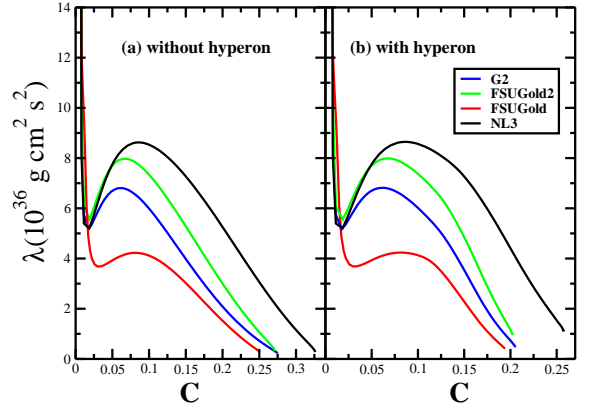}
\caption{(Color online) The tidal deformability $\lambda$ as a function of the
compactness $C$ for the four EOS with and without hyperon. }
\label{fig7}
\end{figure}
\begin{figure}[ht]
\includegraphics[width=1.0\columnwidth]{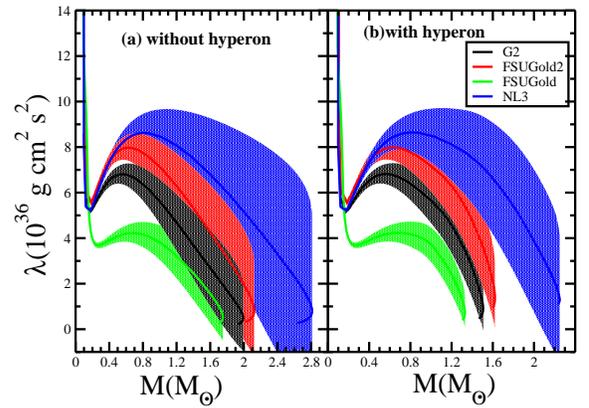}
\caption{(Color online)  Tidal deformability $\lambda$ of a single  neutron 
star as a function of the neutron-star mass for a range of EOSs. 
The estimate of uncertainties in measuring $\lambda$ for equal mass binaries 
at a distance of D = 100 Mpc is shown for the Advanced LIGO detector in 
shaded area. (b) Same as (a), but for hyperon star.}
\label{fig8}
\end{figure}
\begin{figure}[ht]
\includegraphics[width=1.0\columnwidth]{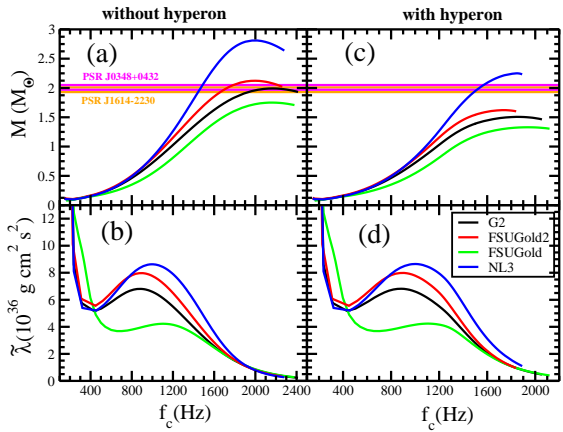}
\caption{(Color online)(a,c) The mass-cut-off frequency f$_{c}$ profile of 
normal and hyperon star using four EOS. (b) The tidal deformabilty with 
cut-off frequency plot of the neutron star(d) Same as (b), but for hyperon 
star. } 
\label{fig9}
\end{figure}
\subsection{Tidal deformability and cut-off frequency of compact star}\label{tidal}
From equation (15), it is known that the tidal deformability $\lambda$ is
a function of the linear tidal love number $k_2$ and the fifth power of the 
radius $R^5$ of the compact star. 
For this purpose, we solve numerically 
Eqns.(12-20) using the initial boundary condition.
To examine the results of tidal deformability with and without hyperons,
we have shown the $\lambda-C$ plot in Fig. \ref{fig7}, where we have 
considered a single neutron star under the influence of an external tidal 
field with adiabatic approximation using the four equations of state. 
In this case, the orbital evaluation time scale is much larger than the 
time scale needed to assume the star as a stationary configuration. 
From the very beginning, we mark an infinitely large $\lambda$ 
corresponding to a small compactness i.e. $C\sim$0.02. Further, the $\lambda$
value falls to a minima that rises again resulting in a hump like pattern
 for each EOS. It is noteworthy that in Fig. \ref{fig7}(b) by introducing
the  NL3 case with hyperon, there is remarkable but mere deviation in $\lambda$ 
value i.e 7.527 g cm$^{2}$ s$^{2}$ ( without hyperon $\lambda$= 7.466 g cm$^{2}$ s$^{2}$). 
 Since, the tidal deformability $\lambda$ is a surface phenomenon, it is very 
much getting affected by the radius of the star in both normal neutron star
and hyperon star. Thus, the tidal deformability  $\lambda$ becomes highly 
sensitive on the radius $R$ even though $k_2$ is small. We estimate the radii 
to be within 12.236$-$14.422 km for a neutron star of  
mass 1.4$M_\odot$ and the range is 13.690$-$14.430 km for neutron-hyperon 
star for all the four stiff or soft equations of state (see Table \ref{tab2}).

Fig. \ref{fig8}, shows the tidal deformability for 
both neutron and hyperon stars. 
We have a large radii  for a smaller stellar mass of $\sim 0.1 M_\odot$ in both cases. At this value of mass and radius, the tidal deformability $\lambda$ becomes 
maximum, because for a large radius with smaller mass, the force of attraction 
within the star is weak and when another star comes closure, the gravitational
pull over ride maximum at the surface part of the star. This phenomena is
true for both neutron as well as hyperon stars \cite{tanja,tanja1}. 
Then, suddenly the tidal deformibility decreases and again 
increases as shown in the figure making a broad peak at around 
M=0.7$-$0.8$M_\odot$ and then decrease smoothly with increase the mass of 
the star. Since, the tidal deformibility depends a lot on both mass and radius 
of a neutron star, it is imperative to measure the radius of the star
precisely, as the mass is already measured with very good precession.
 Recently, Steiner et.al., predicted the most extreme  limit 
for the tidal deformabilities between 0.6 and 6 $\times$ 10$^{36}$ g cm$^{2}$ s$^{2}$
for 1.4$M_\odot$ with 95$\%$ confidence. This range can be constraint on high 
dense matter of any measurements \cite{ste15}. 
Mostly, the binaries masses are about 1.4$M_\odot$, so in particular
we are interested to study the phenomena within this mass range and the
results are summarize in Table \ref{tab2}. 
Comparing the results, we notice that the tidal deformability
$\lambda$ is quite sensitive to the EOS. It is more for stiffer EOS, 
because of the high-density behavior of the symmetry energy \cite{fattev}.

Finally, we calculate the weighted tidal deformability of the binary 
neutron star of mass $m_{1}$ and $m_{2}$ and is approximate is \cite{tanja,tanja1}:
\begin{eqnarray}
\tilde{\lambda} = \frac{1}{26}\left[\frac{m_{1}+12 m_{2}}{m_{1}}\lambda_{1} +
\frac{m_{2}+12 m_{1}}{m_{2}}\lambda_{2} \right], 
\end{eqnarray} 
and the root mean square (rms)  measurement uncertainty 
$\Delta\tilde{\lambda}$ can be calculated following approximate 
formula \cite{tanja,tanja1}:
\begin{eqnarray}
\Delta\tilde{\lambda}\approx \alpha \left(\frac{M}{M_\odot}\right)^{2.5}
\left(\frac{m_{2}}{m_{1}}\right)^{0.1}\left(\frac{f_{cut}}{Hz}\right)^{-2.2}
\left(\frac{D}{100 Mpc}\right), 
\end{eqnarray}
where $\alpha$ = 1.0 $\times$ 10$^{42}$ g cm$^2$ s$^2$ is the tidal deformability
for a single Advanced LIGO detector and $f_{cut}$ ($f_{end}$) cutoff 
frequency \cite{dam} for the end stage of the inspiral binary neutron stars. 
D denotes the luminosity distance from the source to observer.

The weighted tidal deformibility for neutron and hyperon stars
and their corresponding masses as cut-off frequency $f_{cut}$ is shown
in Fig. \ref{fig9}. 
 The cut-off frequency is a stopping criterion to estimate when the tidal
model no longer describes the binary. Here, we take the cut-off to be 
approximately when the two neutron stars come into contact, estimated as in 
Eq.36 of Ref. \cite{dam}. Specifically, we use $f_{cut} = 2 f_{orb.}^{N(R_{1}+
R_{2})}$, where $f_{orb.}^{N(R_{1}+R_{2})}$ is the Newtonian orbital frequency
corresponding to the orbital separation where two unperturbed neutron stars
with radii $R_{1}$ and  $R_{2}$ would touch. 
In the upper panel Fig. \ref{fig9}(a,c), it shows the variation of mass of the 
binary as a function of cut-off frequency $f_{cut}$. Here, we considered $m_1=m_2$,
i.e., both the masses of the binary are equal. 
Initially, the masses of the stars 0.2 $M_\odot$ remain almost constant upto 
$f_{cut}\approx 400$ Hz. Then the mass increases nearly exponentially upto
a maximum mass of $\approx$1.75$-$2.81$M_\odot$ (for NS) and
 $\approx$1.33$-$2.25$M_\odot$ (for hyperon star) and then decreases. By
this time, the cut-off frequency $f_{cut}$ attains quite large value.
When the individual mass of the binary is 1.4$M_\odot$, the NL3 set weighted
tidal deformibility achieve the cut-off frequency $f_{cut}\approx 1256.7$ Hz
is the minimum contrary to the $f_{cut}\approx 1608.0$ Hz of FSUGold 
at the same mass of the single NS. 
It is also clear from the figure that the weighted tidal deformability 
of the NS for the four models are 7.466, 3.668, 5.229 and 2.418 for NL3, G2, 
FSUGold2 and FSUGold, respectively with the corresponding frequency 1256.7,
1440.9, 1332.4 and 1608.0 Hz. 

Using the cut-off frequency, we calculate the uncertainty in the 
measurement of the tidal deformability ($\Delta\tilde{\lambda}$) 
obtained from these four EOSs for an equal-mass binary star inspiral 
at 100 Mpc from aLIGO detector (shaded region in Fig. \ref{fig8}). 
The uncertainty in the lower mass region (0.4$-$1.0$M_\odot$) 
of the NS $\Delta\tilde{\lambda}$ is smaller. Similar results are found
in the case of hyperon star also.
Interestingly, the error ($\Delta\tilde{\lambda}$) increases with 
increase the mass of the binary for all the EOSs. 
From Table \ref{tab2}, by comparing the $\Delta\tilde{\lambda}$ obtained from all
the EOSs, we find that predicted errors are greater than the measured value
for a star of mass 1.4$M_\odot$.

\section{Summary and Conclusions}\label{conc}
In summary,  four different models have been extensively applied which are
obtained from effective field theory motivated relativistic mean field 
formalism. This effective interaction model satisfies the nuclear saturation 
properties and reproduce
the bulk properties of finite nuclei with a very good accuracy. We used these
four forces of interaction and calculate the equations of state for neutron
and hyperon stars matter. It is  noteworthy that each term of the interaction
has its own meaning and has specific character. The inclusion of extra 
terms (nucleons replaced by baryons octet) in the Lagrangian contribute to 
soften the EOS and the matter becomes less compressible. Hence, there is 
decrease in the maximum mass by $\sim 0.5M_{\odot}$ than the pure neutron star. 

We have extended our calculations to various tidal responses both
for electric-type (even-parity) and magnetic-type (odd-parity) of 
neutron and hyperon stars in the influence of an external gravitational 
tidal field. The love numbers are directly connected with surficial love 
number $h_{l}$ associated with the surface properties of the stars. 
Subsequently, we study the quadrupolar tidal deformability $\lambda$ 
of normal neutron star and hyperon star using different set of equations
of state. These tidal deformabilities particularly depend on the 
quadrupole love number $k_{2}$ and radius ($R$) of the isolated star. 
Although the maximum value
of $k_{2}$ is not very sensitive to the EOS for neutron and hyperon stars
lying in the range $k_{2}\approx 0.144-0.170$ and $0.143-0.194$ for
neutron and hyperon stars, respectively, but it is very much sensitive 
to the radius of the star. 

We find that aLIGO can constraint  on the existence of hyperon star,
i.e., the inner core of the NS has hyperons, but detecting them
can be much harder. 
However, it should be able to constraint the neutron 
star deformability to $\lambda\leq$10 $\times$ 10$^{36}$ g cm$^{2}$ s$^{2}$ for a 
binary of 1.4$M_\odot$ neutron stars at a distance 100 Mpc from the detector. 
Also, the present calculations suggest to use the portion of the signal 
with the gravitational wave frequency less than 400 Hz. 
In future, we expect that aLIGO should be able to measure $\lambda$ even for
neutron stars masses up to 2.0$M_\odot$ and consequently  constraint the
stiffness of the equations of state.\\

{\bf ACKNOWLEDGEMENTS:}

Bharat Kumar would like to take this oppertunity to convey special thanks Tanja Hinderer whose keen interest, fruitful discussions and useful suggestions.

\end{document}